\documentclass[conference]{IEEEtran}
\IEEEoverridecommandlockouts
\usepackage{cite}
\usepackage{amsmath,amssymb,amsfonts}
\usepackage{algorithm,algorithmic}
\usepackage{newtxmath}
\usepackage{graphicx}
\usepackage{textcomp}
\usepackage{xcolor}
\usepackage{subcaption}
\usepackage[hyphens]{url}
\def\BibTeX{{\rm B\kern-.05em{\sc i\kern-.025em b}\kern-.08em
    T\kern-.1667em\lower.7ex\hbox{E}\kern-.125emX}}
\begin{document}

\title{Topiary: Fast, Scalable Publish/Subscribe for Peer-to-Peer (D)Apps}

 \author{\IEEEauthorblockN{1\textsuperscript{st} Yifan Mao}
 \IEEEauthorblockA{\textit{Department of
Computer Science and Engineering} \\
 \textit{The Ohio State University}\\
 Columbus, United States \\
 mao.360@osu.edu}
 \and
 \IEEEauthorblockN{2\textsuperscript{nd} Shaileshh Bojja Venkatakrishnan}
 \IEEEauthorblockA{\textit{Department of
Computer Science and Engineering} \\
 \textit{The Ohio State University}\\
 Columbus, United States \\
 bojjavenkatakrishnan.2@osu.edu}
 }

\maketitle

\begin{abstract}
The emergence of blockchain technology has fostered the development of numerous decentralized applications (dapps) in recent years
Pub/sub (publish/subscribe) systems play a crucial role by associating messages with specific topics and propagating them from publishers to subscribers across the network.
Decentralized pub/sub aims to provide this functionality without relying on centralized control or global network state information, enabling message propagation among nodes in a coordinated manner. Efficiency in pub/sub services entails ensuring that subscribers receive published messages promptly.
We introduce Topiary, a rapid and scalable protocol designed for decentralized applications' pub/sub systems. Topiary autonomously learns an efficient peer-to-peer (p2p) topology tailored to the publish/subscribe network. It does so by analyzing peers' interactions with their neighbors. Inspired by concepts from the multi-armed bandit problem, Topiary strikes an optimal balance between maintaining connections with well-connected neighbors and exploring new connections within the network, based on their topical needs.
Through experimental evaluations, Topiary has shown a 50\% reduction in broadcast latency while achieving an interested topic coverage of over 98\%, marking it as a promising solution for efficient decentralized pub/sub networks.
\end{abstract}


\section{Introduction}
\label{s: introduction}
The exponential growth of blockchain technology has paved the way for numerous decentralized applications (dapps) across various sectors.  
These dapps often operate on a network of independent servers connected through a peer-to-peer architecture, catering to client devices. 
They can be implemented in different ways—leveraging smart contracts on blockchains like Ethereum (e.g., ERC 20 tokens), existing within a blockchain (e.g., Filecoin, Livepeer), or even without using a blockchain (e.g., Mastodon, IPFS). These applications span a wide array of domains such as finance, gaming, healthcare, social networking, and more.
Today dapps abound in diverse domains spanning payments, decentralized finance, gaming, healthcare, social networks etc.
A fundamental messaging primitive used in many dapps is publish/subscribe or pub/sub for short.  
In a pub/sub system, each message is associated with a distinct topic; each node has a preference for receiving messages that belong to one or more topics, which the node indicates by ``subscribing" to the topics.
Messages in a topic can be published by any node subscribed to the topic in the network.  
A node that is subscribed to a topic must receive all messages tagged under that topic. 
Pub/Sub is used routinely in centralized cloud systems, and is beginning to see widespread adoption in large-scale dapps.  
E.g., in the Streamr~\cite{savolainen2020streamr} dapp, clients can subscribe to real-time datastreams they are interested in, which are then delivered through a network of brokers using pub/sub. 
The Stremr protocol can be used to implement applications such as chat rooms (in which each room is a topic), collaborative document editing etc. 
In Eth2.0, nodes can subscribe to topics such as `beacon\_block', `proposer\_slashing' etc. for the consensus protocol operation. 
Pub/Sub is also critical for  implementing data availability sampling in which blocks are erasure coded and  coded chunks  must be distributed to nodes in the network~\cite{al2018fraud}.   

The examples outlined above are global-scale systems encompassing several thousands of nodes, potentially spanning multiple continents and autonomous systems. 
In decentralized pub/sub, we seek to achieve the pub/sub functionality without relying on any centralized control for knowing the global network state and coordinating different nodes.  
For efficient application performance, it is desirable that the published messages are received by the subscribers as quickly as possible with minimal delay.   
E.g., in Eth2.0 nodes must receive consensus-related messages within the slot deadline of 12 seconds. 
Low latency is also important in applications such as online chat. 
A na\"ive approach to realizing pub/sub is simply broadcasting all messages over the entire network. 
However, under this approach nodes would also receive messages in topics they are not interested in which leads to wasted network resources. 

To improve network efficiency, a common practice is to construct separate overlays for each topic, thereby restricting broadcast of a message in a topic to be within the topic's overlay.  
Indeed, the popular GossipSub~\cite{vyzovitis2020gossipsub} implementation (which is used by Eth2.0 among other dapps) uses this approach. 
However, if the number of topics in the dapp is large (e.g., $>$1000 chat rooms in the chat dapp), and if a node is interested in a large number of topics, then the node must be part of a large number of separate overlays in GossipSub. 
Simultaneously connecting to a large number of overlays requires the node to create and maintain a large number of TCP connections, which is prohibitive.   

In this paper, we consider the problem of how to construct an efficient pub/sub topology to minimize message latency and maximize the interested topic coverage in wide-area p2p networks especially with large size of topics.
Crucially, we seek constructions in which the total number of connections a node must make is bounded even if the number of topics is large. 
If there is just one topic (i.e., all messages are broadcast to all nodes), prior works~\cite{mao2020perigee, bowengoldfish} have shown that constructing a topology by considering the relative location of peers in the network can lead to significant speedup in broadcast delay compared to a random topology. 
If two peers share interests in several topics, then using a single connection link between the peers, messages can be exchanged on any of those shared topics. 
However, constructing an efficient topology that considers not only the relative location of peers in the network, but also their topic preferences to minimize the number of connections is a highly non-trivial problem. 




We present Topiary, a decentralized protocol for constructing an efficient topology in a p2p pub/sub network.
In Topiary, we propose a data-driven algorithm in which a pub/sub node adaptively decides which neighbors it should connect to, purely based on the node's past interactions with its neighbors and the message topics for whether the receiver is interested. 
Our protocol is motivated by the classical multi-armed bandit problem~\cite{auer2002using}: we view the choice of neighbors a peer connects with as a (combinatorial) arm and the pub/sub latency performance as the reward for choosing the arm. 
Following an exploration-exploitation paradigm for neighbor selection in a p2p network has been previously proposed in the literature~\cite{mao2020perigee,babel2022strategic,math11234741}. 
Our present proposal differs from prior work in two aspects. 
First, in Topiary a node balances retaining old neighbors and exploring new neighbors over multiple topics while prior work has considered networks with just a single topic. 
Second, Topiary proposes a novel peer exploration strategy that is designed to minimize the number of overall connections required. 

Topiary proceeds in rounds in which each round a node observes its interactions with its neighbors by looking at the messages' arrival times. 
Based on the observed performance, the node decides which set of neighbors to connect with for the next round by favoring neighbors that forward relevant messages fast or neighbors that share significant topic interests with the node.
In practice, a round can be defined to span a few minutes or seconds depending on the frequency of message publication in the topics. 
The performance of any subset of neighbors is evaluated by measuring the relative quickness with which the subset of neighbors relay messages compared to other subsets of neighbors and by estimating whether all messages in the topic are being relayed. 
At the beginning of a round, a node retains the subset of neighbors from the previous round that has the best performance and makes connections to a small number of new (potentially previously unseen) neighbors.
Our approach of purely using past observed performance (promptness of message arrival, message loss) to select future neighbors is robust to various heterogeneity in the network such as differences in computation power of nodes, network bandwidth, peer locations among others.  
Our approach also allows a user to specify importance weights to the different topics, through which the constructured topology can be biased to favor certain topics more than others. 
Through experiments we show that Topiary forms efficient topologies under various settings: overall Topiary improves message propagation delay by 50\% and decreases message loss to less than 5\% compared to the random connections used in GossipSub~\cite{vyzovitis2020gossipsub}.

Prior works have considered modifying the p2p topology for faster message propagation in decentralized pub/sub networks.
Gossipsub~\cite{vyzovitis2020gossipsub} proposes a pub/sub protocol to score the neighbors to incorporate resilience against a wide spectrum of attacks.
However, such topology only distributes the neighbor selection to the limited interested topic and can hardly deal with the network that with less neighbor connection degree but with a high demand of topics.
They are also oblivious to link latencies between the honest nodes that they have not made full use of information from honest neighbors as they focus more on avoiding attacks, which renders their performance to be slightly better than the random topology but has a gap to the ideal connection topology.
They cluster the nodes by their interested topics, this can help to ensure the messages reach the most interesting subscribers with lower latency.
However, such group method is impossible with various topics, and the group connection is not efficient enough.
Complete connections in each group cost huge bandwidth, and random connections in each group can hardly cover all the topics that each node is interested in.

In contrast, our algorithm does not group the nodes into each cluster but score the current neighbor subset in a union and score the future connected nodes individually to make up the union, which can make the best of the nodes' information to decrease the propagation latency.

\section{Model}
\label{s: model}

\subsection{Network Model}
\label{s:netmodel}
We model the pub/sub network as an undirected graph $G(V, E)$ comprising of a set of nodes $V$ and communication links $E$ between them.\footnote{We use the terms node and peer interchangeably.} 
A node can make outgoing connection requests to at most $d$ peers and accept any number of incoming connection requests from other peers.
Once a connection is established, data can flow bidirectionally over the link. 
A node can specify the set of neighbors to which to connect, to optimize performance. 
$\Theta$ is the set of topics in the network. 
For a node $v \in V$, let $\sigma_v \subseteq \Theta$ be the set of topics $v$ is interested in (i.e., subscribes to). 

Time proceeds in discrete rounds, $t = 0, 1, 2, ..$. 
Each round a publisher from each topic publishes a message, which is then gossiped over the network following a gossip protocol.
Each published message belongs to exactly one topic. 
For a message $m$, we let $\theta(m)$ be the topic the message belongs to. 
The publisher for topic $\theta \in \Theta$ at time $t$ is chosen uniformly at random from the set $\{ v \in V: \theta \in \sigma_v \}$ of nodes interested in topic $\theta$. 
This randomness is independent across time and across topics. 

When a publisher generates a new message, it gossips the message to all its neighbors in $G$.
Each link $(u, v) \in E$ has a latency $l_{(u,v)} \geq 0$ which is the time it takes for the message to propagate from $u$ to $v$ (we assume $l_{(u,v)} = l_{(v,u)}$) through the link $(u, v)$. 
Links $(u, v) \notin E$ also have an associated $l_{(u,v)}$ which is the time it takes for a message to propagate from $u$ to $v$ had there been a direct connection between $u$ and $v$. 
We think of $l_{(u,v)}$ as the latency induced by the locations of nodes $u$ and $v$ on the physical network. 
When a node $v$ receives a new message from its neighbor $u$, it checks the message header for the topic information and decides whether to forward the message to its other neighbors.
We let $\delta_v > 0$ be the time it takes for node $v$ to process 
the message header. 
If a node receives a message that it has previously seen, it does not forward the message to its neighbors.

Each message includes a $TTL \in \{0, 1, 2, \ldots \}$ (time to live) field in its header. 
Publishers choose an initial value for the $TTL$ when they publish a message. 
The $TTL$ value of a message is decremented by one whenever a node that is not interested in the message's topic receives the message.  
The $TTL$ value of a message is unchanged if a node that is subscribed to the message's topic receives the message.
A node relays a message with a positive $TTL$ to all its neighbors, except  the one from which it received the message. 
If the $TTL$ of a message is zero, a node relays the message only to neighbors that are interested in the message's topic. 


We assume each node knows the IP addresses and subscription information of all other nodes in the network. 
Thus, a node can send connection requests to any other node in the network. 
However, a node knows link latency information only for the links with its neighbors.  
A node does not know the node processing delays of other nodes. 


\subsection{Problem Statement}

Under the network model presented in Section~\ref{s:netmodel}, our goal is to design a decentralized algorithm for constructing a topology of the p2p pub/sub network such that, 
\begin{enumerate}
\item messages published in a topic are received by as many interested nodes as possible; ideally all nodes subscribed to the topic. 
Note that that latter is satisfied if nodes subscribed to a topic form a connected subgraph. 
We call this property as topic connectivity. 
\item messages published in a topic are received by subscribers with as low a latency as possible. 
\item the amount of bandwidth nodes waste in receiving and forwarding messages of unsubscribed topics is minimized.  
\end{enumerate}
The objectives listed above may not be simultaneous optimized in all problem instances. 
Hence, we consider optimizing a weighted combination of the objectives. 

\section{Motivation}
\label{s:motivation} 
\begin{figure}
  \centering
  \begin{subfigure}{0.22\textwidth}
    \includegraphics[width=\textwidth]{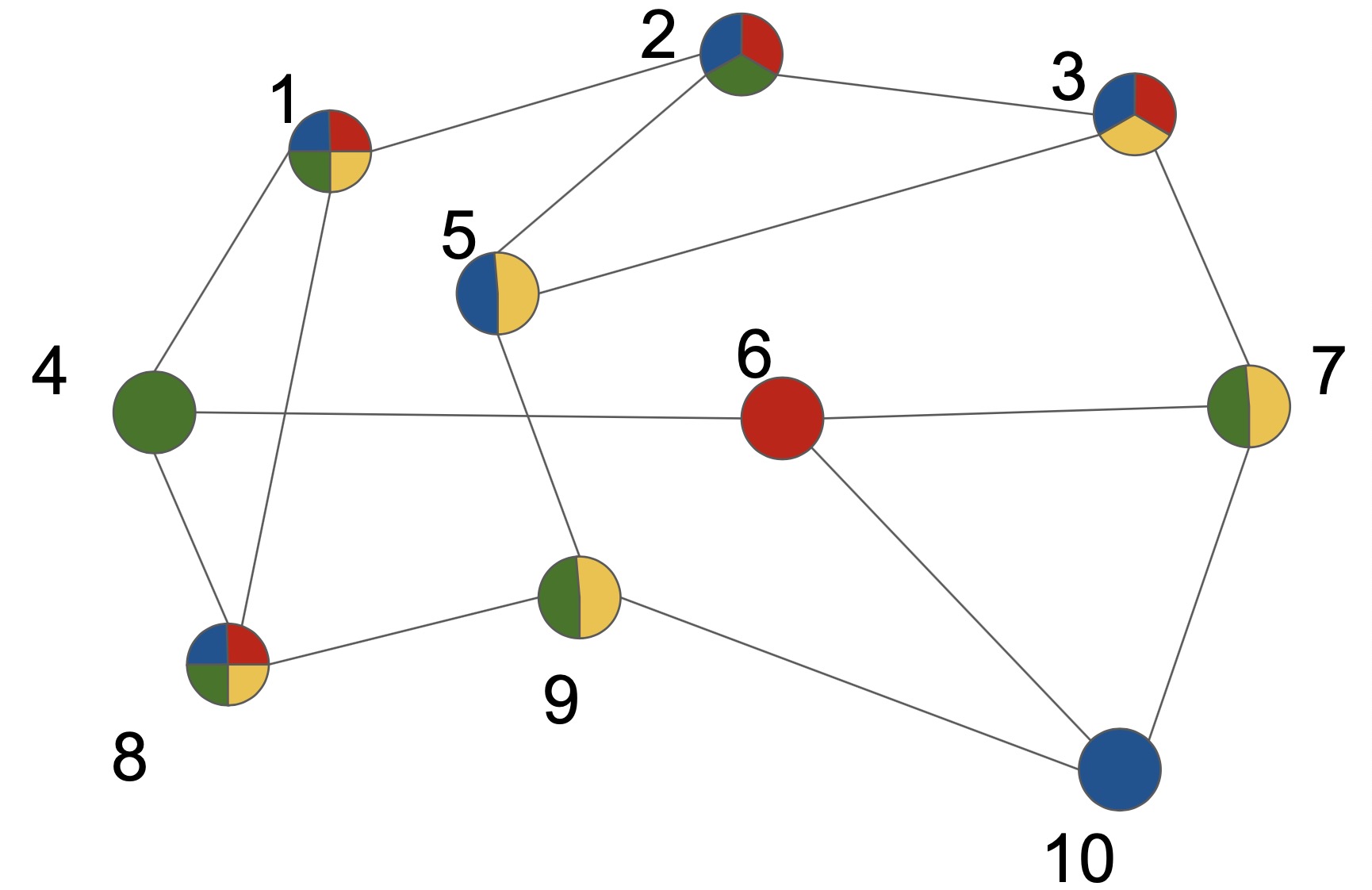}
    \caption{Random} 
    \label{fig:moti-random}
  \end{subfigure}%
  \hspace*{\fill}
  \begin{subfigure}{0.22\textwidth}
    \includegraphics[width=\textwidth]{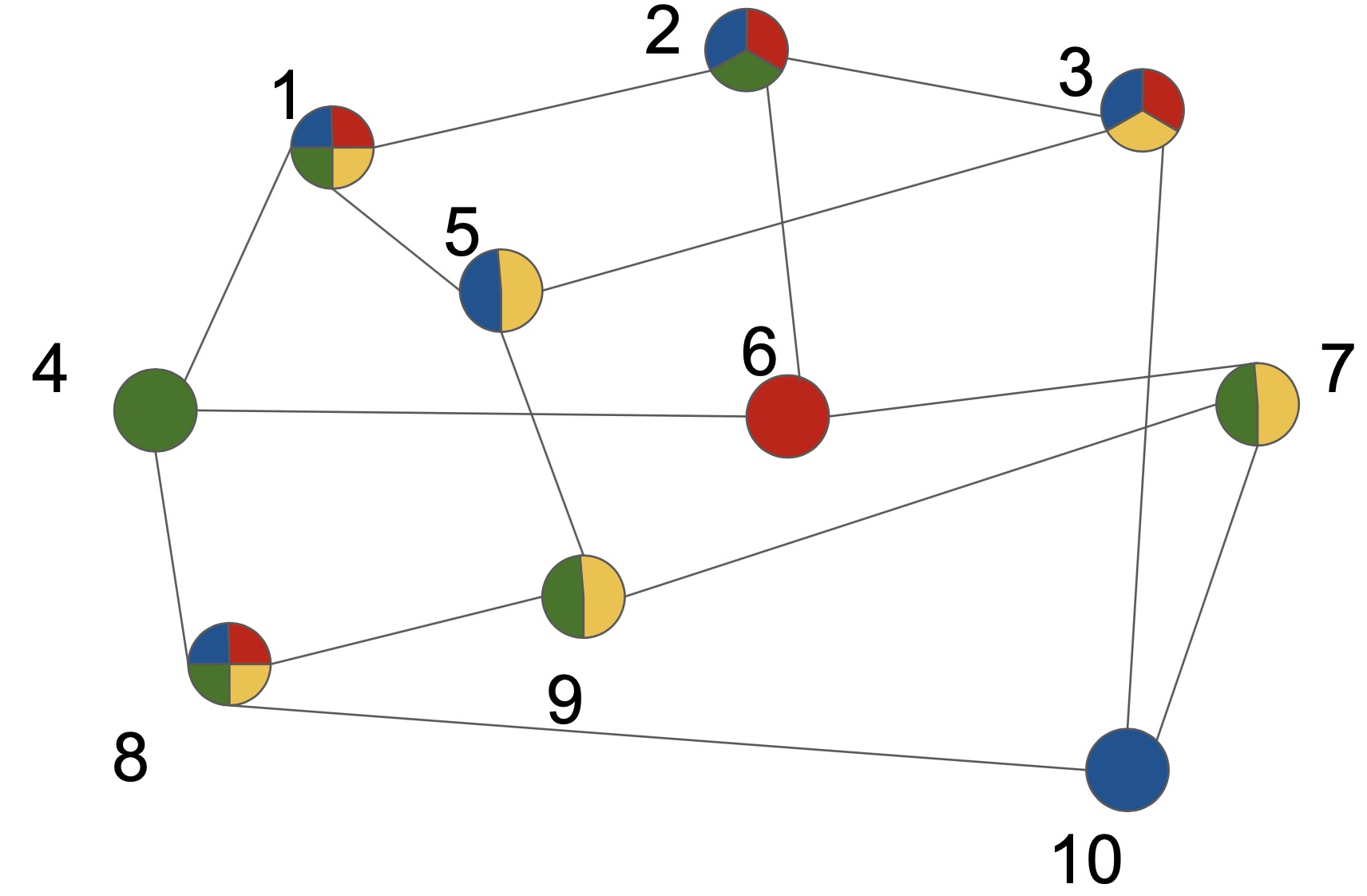}
    \caption{Subgraph} 
    \label{fig:moti-policy}
  \end{subfigure}%
  \caption{Example of a 10-node network topology used in  (a) random, and (b) interest subgraph. Nodes here may be interested in 4 topics in \{red, yellow, green, blue\}.}
\label{fig:moti-topology}
\end{figure}

We motivate our results using a simple toy example, that illustrates the effect of network topology on pub/sub efficiency. 
We consider a network of 10 nodes with a degree of 3.
There are four topics: \{red, yellow, green, blue\}.
Fig.~\ref{fig:moti-topology} shows the topic interests for the 10 nodes. 
E.g., node 1 is subscribed to topics \{green, red, blue\}, node 2 is subscribed to all four topics, and so on. 
All links $(u,v)$ have a delay $l_{(u,v)} = 1$. 

We first consider a random topology of degree 3 in Figure~\ref{fig:moti-random}.
Due to the random connections, nodes interested in a topic do not necessarily form a connected subgraph. 
E.g., node 5 does not have any neighbors interested in the red topic, and is thus disconnected from the red topic subgraph. 
Similarly, node 9 is disconnected from other subscribers of the blue topic.
Thus, if messages are to reach all interested subscribers, we require a sub-optimal initial $TTL$ value of one resulting in  network bandwidth wastage. 

On the other hand, if we carefully build a topology based on nodes' topic preferences we can achieve connected topic subgraphs with the same degree as shown in Figure~\ref{fig:moti-policy}. 
Each node has at least one neighbor that shares an interest in each topic the node subscribes to.
Thus, even with an initial $TTL$ of zero, we can ensure messages of a topic reach all subscribers of the topic.

\begin{figure}[!tbp]
  \centering
  \includegraphics[width=0.42\textwidth]{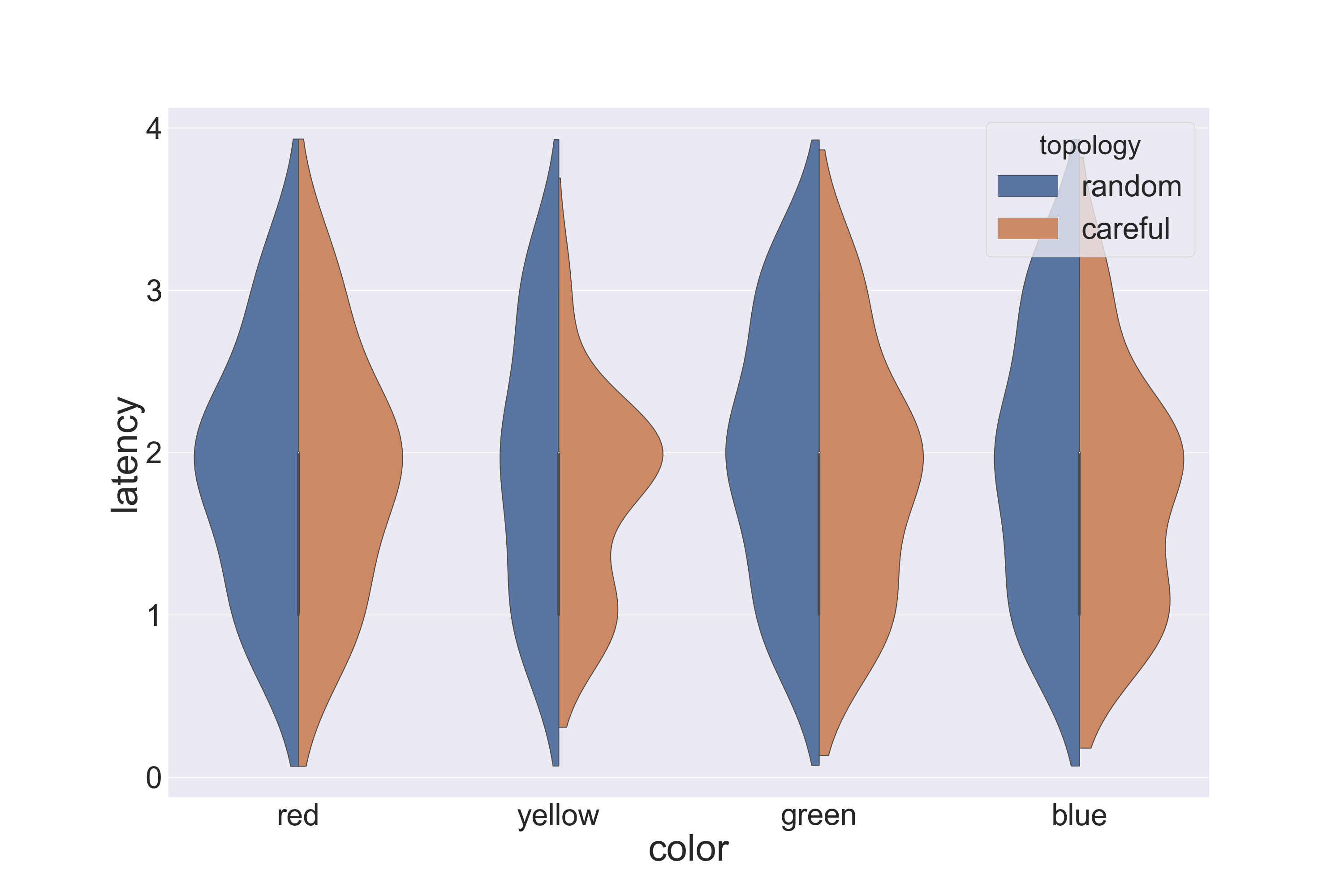}
  \caption{Latency distributions for the topologies in Figure~\ref{fig:moti-topology} on different topics. }
  \label{fig:moti-degree}
\end{figure}

To quantify the performance differences between the two topologies discussed above, in Figure~\ref{fig:moti-degree} we plot histograms of the shortest path distances between all pairs of nodes interested in a topic. 
The carefully constructed  topology has a lower average latency of 1.8, while the random topology has an average latency of 1.96.

The observation illustrates that a good topology can optimize the pub/sub network, to decrease message propagation latency and increase topic subgraph connectivity.
Different from the prior works that focus on lowering latency for just one topic~\cite{mao2020perigee,bowengoldfish}, 
in a pub/sub network the algorithm must  take all topics into consideration and optimize both message delivery latency and topic connectivity under possibly heterogeneous network bandwidth, peer location and compute capabilities. 

\section{Algorithm}
\label{sec:algorithm}
Topiary is a fully decentralized algorithm designed for constructing topologies within large-scale, wide-area peer-to-peer (p2p) pub/sub networks. 
The algorithm draws loose inspiration from the combinatorial multi-armed bandit (MAB) problem \cite{gai2010learning,kveton2015tight,chen2013combinatorial, dong2023graph}, treating each node as a MAB agent and considering the set of peers it selects to connect with as the combinatorial MAB action. 
The reward for taking an action is determined by how efficiently a node receives messages and the amount of bandwidth used in relaying messages related to uninterested topics. 
In Topiary, peers operate independently without the need for trust or cooperation with other peers. Decisions are solely based on a peer's observations of previously connected peers and messages received. 
This characteristic makes Topiary highly adaptable to network dynamism and heterogeneity. Moreover, Topiary boasts security measures against Sybil and Eclipse attacks.

Algorithm~\ref{alg:total algorithm} outlines the primary steps executed by Topiary at a node $v$. 
An 'epoch' refers to a fixed number of rounds (in our experiments, we set an epoch to equal 1000 rounds). 
Throughout an epoch, node $v$ gathers information concerning its current set of neighbors' message relay speed across various topics of interest. 
Additionally, the node estimates whether it received all published messages from subscribed topics or experienced message losses. 
Accurately estimating missed messages depends on the specific application and falls beyond the paper's current scope. Subsequently, the node calculates the amount of bandwidth spent on downloading irrelevant messages unrelated to its subscribed topics.

Utilizing the gathered information, $v$ computes a 'score' for different neighbor subsets, considering subsets of approximately 60\% of a node's set degree $d$ for scoring purposes. 
Connections to the subset nodes with the highest scores are retained for the subsequent epoch, while connections to other nodes are terminated. 
To fill the vacancies left by terminated connections, node $v$ establishes connections with randomly selected peers according to an exploration policy. 
Below, we delve into the computation of scores and the process for randomly selecting peers for exploration.

\begin{algorithm}[t]
\caption{Topiary: algorithm template for node $v$}
\label{alg:total algorithm}
\begin{algorithmic}[1]
\STATE {\bf Input:} set of neighbors $\Gamma_v$ of node $v$ in current epoch, topic subscriptions $\sigma_v$, degree bound $d_v$
\STATE {\bf Output:} set of peers to connect with during next epoch. 
\STATE // Collect observations of neighbors 
\STATE $\mathcal{O}_v \leftarrow \{ \}$
\FOR{each message $m$ received during epoch} 
\STATE $\mathcal{O}_v \leftarrow \mathcal{O}_v \cup $ neighbors' timestamp metadata about $m$ 
\ENDFOR 
\STATE // Estimate fraction of missing messages 
\STATE $f_v \leftarrow \{ \} $
\FOR{each topic $\theta \in \sigma_v$}
\STATE $f_{v, \theta} \leftarrow$ fraction of messages on topic $\theta$ not received by $v$ during epoch 
\STATE $f_v \leftarrow f_v \cup \{ f_{v, \theta} \}$
\ENDFOR
\STATE // Compute score for neighbor subsets 
\FOR{$\Gamma \subset \Gamma_v$ such that $|\Gamma| = 0.6 d$ }
\STATE Compute score $S(\Gamma) \leftarrow$ \textsc{Score}$(\Gamma, \mathcal{O}_v, f_v)$
\ENDFOR
\STATE // Compute best performing subset of neighbors
\STATE $\Gamma^* \leftarrow \text{argmax}_{\Gamma \subset \Gamma_v, |\Gamma| = 0.6 d} S(\Gamma)$
\STATE $\Gamma^\mathrm{explore} \leftarrow $ \textsc{Explore}$(\mathcal{O}_v, f_v)$
\STATE return $\Gamma^* \cup \Gamma^\mathrm{explore}$
\end{algorithmic}
\end{algorithm}

\subsection{Topic delay score}
\label{sec:reward model}

Let $M_v$ be the set of messages received by node $v$ during an epoch. 
For each message $m \in M_v$, $v$ records the local timestamp $T^m_u$ when each neighbor $u \in \Gamma_v$ sent $m$ to $v$ ($T^m_u = \infty$ if  $u$ did not send $m$ to $v$). 
$\Gamma_v$ here denotes the set of neighbors of $v$ during the epoch. 
Given the impracticality of achieving perfectly synchronized clocks across nodes in a decentralized network, we rely solely on node $v$'s local clock to estimate the speed at which different neighbors relay messages.
To do this, let $T^m_\mathrm{min} = \min_{u \in\Gamma }T^m_u$ be the timestamp of when $v$ received $m$ for the first time. 
We normalize the collected absolute timestamps as follows: 
\begin{align}
\tilde{T}^m_u = T^m_u - T^m_\mathrm{min}, \forall u \in \Gamma_v.  
\end{align}
Let $M_v^\Gamma = \{m \in M_v: \min_{u \in \Gamma} \tilde{T}_u^m < \infty \}$ be the subset of messages that are delivered to $v$ by at least one node in $\Gamma$. 
The normalized timestamps $\{\tilde{T}^m_u: u \in \Gamma_v, m \in M_v \}$ are used to compute a topic delay score for each subset $\Gamma \subset \Gamma_v$ as
\begin{align}
F_d(\Gamma) = \sum_{m \in M^\Gamma_v: \theta(m) \in \sigma_v} \frac{\min_{u \in \Gamma} \tilde{T}^m_u}{ |\{m: m \in M^\Gamma_v, \theta(m) \in \sigma_v \}|}.  
\end{align}
The topic delay score for a subset $\Gamma$ captures the average relative delay the subset takes to deliver messages compared to other neighbors. 

    

\subsection{Topic coverage score}
For each topic $\theta \in \sigma_v $, let $E_\theta$ represent an estimate of the number of messages published on topic $\theta$ during the epoch. 
However, the method of computing such an estimate is application-specific and beyond the scope of this discussion. 
For instance, a node might utilize prior knowledge concerning the frequency of message publications for this estimation. 
Alternatively, an estimate could be derived by utilizing sequence numbers present in message headers. 
The topic coverage score for each subset $\Gamma \subset \Gamma_v$ quantifies the fraction of messages that $\Gamma$ fails to relay and is calculated as:
\begin{align}
F_c(\Gamma) = \frac{\sum_{m \in M_v: \theta(m) \in \sigma_v }  \mathbf{1}_{ \min_{u \in \Gamma} \tilde{T}^m_u < \infty}  }{\sum_{\theta \in \sigma_v}E_\theta}. 
\end{align}
It's important to note that the topic delay score $F_d(\Gamma)$ measures the average latency for messages delivered by $\Gamma$ exclusively. 
On the other hand, the topic coverage score $F_c(\Gamma)$ encapsulates the complementary aspect, representing messages that were not delivered by $\Gamma$.


\subsection{Bandwidth wastage score}

Lastly, we calculate a score representing the bandwidth wastage resulting from receiving messages on topics that are not of interest to node $v$ from a specific subset of neighbors $\Gamma \subset \Gamma_v$, denoted as:
\begin{align}
F_w(\Gamma) = \sum_{m \in M_v: \theta(m) \notin \sigma_v}\mathbf{1}_{\min_{u \in \Gamma} \tilde{T}^m_u < \infty }. 
\end{align}



\subsection{Overall score }
\label{sec:score function}
The overall score for a subset $\Gamma \subset \Gamma_v$ is computed as 
\begin{align}
S(\Gamma) = w_c F_c(\Gamma) + w_d F_d(\Gamma) + w_w F_w(\Gamma),  \label{eq:overall score}
\end{align}
Here, $w_c > 0$, $w_d > 0$, and $w_w > 0$ are user-defined parameters that determine the weight allocated to each term in the optimization objective. 
For instance, in scenarios where data loss is intolerable, a prioritization can be set such as $w_c > w_d > w_w$. 
It's important to note that in this context, a higher score indicates poorer performance. Equation~\eqref{eq:overall score} represents the \textsc{Score} function outlined in Algorithm~\ref{alg:total algorithm}.

At the end of the epoch, node $v$ retains its connections to nodes within the subset $\Gamma^*$ that exhibits the highest score among all subsets, each of which constitutes 60\% of the node's degree bound $|\Gamma_v|$. This determination is made using the computation derived from Equation~\eqref{eq:overall score}.



\subsection{Exploration}
\label{sec:exploration}
Following the retention of the subset with the highest score, node $v$ proceeds to select replacement nodes for the connections that were terminated. 
Our objective is to select new neighbors capable of complementing any suboptimal performance exhibited by the retained subset $\Gamma^*$ on specific topics, if such shortcomings exist. Simultaneously, we aim to explore uncharted neighbors that might offer superior performance compared to the current connections. 
To balance these objectives, we assign weights to nodes and conduct random sampling based on these weights.

First, we identify the topic(s) for which the retained subset $\Gamma^*$ has suboptimal performance. 
For each topic $\theta \in \sigma_v$ we compute a score 
\begin{align}
S^\theta(\Gamma^*) = w_c F_c^\theta(\Gamma^*) + w_d F_d^\theta(\Gamma^*), \label{eq: per topic score}
\end{align}
where 
\begin{align}
F_d^\theta(\Gamma^*) &= \sum_{m \in M^{\Gamma^*}_v: \theta(m) = \theta} \frac{\min_{u \in \Gamma^*} \tilde{T}^m_u}{ |\{m: m \in M^{\Gamma^*}_v, \theta(m) = \theta \}|}    \\ 
F_c^\theta(\Gamma^*) &= \frac{\sum_{m \in M_v: \theta(m) = \theta }  \mathbf{1}_{ \min_{u \in \Gamma^*} \tilde{T}^m_u < \infty}  }{E_\theta}.
\end{align}
The score in Equation~\eqref{eq: per topic score} is analogous to the score in Equation~\eqref{eq:overall score}, except that the topic delay and topic coverage scores are computed over messages of a specified topic $\theta$ instead of all topics.
We then identify the topic having the best score, $\min_{\theta \in \sigma_v} S^\theta(\Gamma^*)$, and define 
\begin{align}
\sigma_v^+ = \{ \theta \in \sigma_v : S^\theta(\Gamma^*) > \eta \min_{\theta \in \sigma_v} S^\theta(\Gamma^*) \},
\end{align}
to be the topics that $\Gamma^*$ does not perform as well on. 
$\eta > 1$ is a thresholding parameter. 

Next, for each node $u \in V, u \notin \Gamma^*, u \neq v$, we set a weight 
\begin{align}
    \omega_u = |\sigma_u \cap \sigma_v^+|.  
\end{align}
The replacement nodes are then sampled (without replacement) from the set $V \backslash (\Gamma^* \cup \{ v\})$ with the probability of sampling a node $u$ proportional to $\omega_u$. 
In scenarios where the subset $\Gamma^*$ exhibits subpar performance across several topics, a basic strategy for exploration might involve locating a set of neighbors. 
Each neighbor would be subscribed to one of the underperforming topics, ensuring that collectively, the neighbors cover all these topics. 
However, this approach could lead to an excessive number of neighbors, surpassing the specified degree bound $d$ for node $v$. 
To mitigate the requirement for a large number of new neighbors, our proposed approach prioritizes neighbors subscribed to multiple of the underperforming topics.



\section{Evaluation}
\label{s: evaluation}
In this section, we discuss the implementation of the algorithm and evaluate the achieved results.

\subsection{Experiment Setup}
\label{s: setup}

\smallskip
\noindent
\textbf{Network model.} 
We consider two network models in our experiments.
\begin{itemize}
\item 
In the first model, we have 1000 nodes that are embedded at random locations within a unit square. 
The link delay $l_{u,v}$ between any two nodes $u, v$ is the Euclidean distance between the nodes on the unit square. 
Each node selects $d = 6$  outgoing neighbors and accept all incoming connection invitations.
\item In the second network model, we have 246 nodes located at cities around the world. 
In this case, the latency $l_{u,v}$ between two nodes $u, v$ is obtained from a dataset on global ping statistics over the Internet (collected on July 19th \& 20th, 2020 from WonderNetwork~\cite{wondernetwork})  
Nodes in the second network select $d = 5$  outgoing neighbors due to the smaller network size, and accept all incoming connection requests.
\end{itemize}

The distinction between incoming and outgoing connections arises only during connection establishment.
Once connections have been established a node receives and relays data on both its incoming and outgoing connections. 

\smallskip
\noindent
{\bf Baselines.}
We compare Topiary against the following algorithms as baselines: 
\begin{itemize}
\item {\em Random graph.} Nodes randomly pick the neighbors with the amount of their outgoing degree.
\item {\em Complete graph.} All the nodes have direct connections to all other nodes. A complete graph is an unrealistic topology, but serves as a reasonable lower bound for performance. 
\item {\em GossipSub.} Nodes equally distribute their outgoing degree to each of their interested nodes to enable at least one connections to that topic.
\item {\em Scribe.} Nodes construct groups and build tree structures inside each group. Here we consider grouping with topic or without topic information.
\end{itemize}

\smallskip
\noindent
{\bf Topics and subscriptions.}
To decide node subscriptions, we use a topic interest rate parameter.
A topic interest rate of $x$\% means each node randomly subscribes to each topic with a probability of $x/100$ (randomness is independent across topics and nodes). 
Therefore, each topic has on average $xn/100$ subscribers where $n$ is the number of nodes in the network. 
To ensure nodes have at least one interested topic and there are overlapping topics interests between nodes, we do not consider the case of small topic interest rates under a small topic size.

\smallskip
\noindent
\textbf{Publishing.}
The network randomly decides the publishers for each topic to generate new messages for that topic (Section~\ref{s:netmodel}).
The role as publisher or subscriber to topics are independent to each other.

\smallskip
\noindent
\textbf{Time to live.}
Each message has a Time-to-Live ($TTL$) value in its header.
We set a default $TTL$ value to 1 at the beginning of the experiment, so each propagation allows one uninterested node to forward this message in the propagation path. 

\smallskip
\noindent 
{\bf Epochs.}
We split the experiments into epochs, each epoch contains 1000 new messages.
Publishers generate new messages independently, other nodes receive and forward the messages by the gossiping algorithm to all their neighbors.
Nodes score their neighbors during the epoch and make the neighbor selection decision at the end of each epoch.
Each node can switch 2 neighbors each time, so they select the best 4-node-subset in the 1000-node network or 3-node-subset in the 246-node network.
We run the experiment for 150 epochs and record all the timestamps that nodes receive their interested topic messages.

\smallskip
\noindent
{\bf Performance measures.}
We show the results by measuring the following parameters:
\begin{enumerate}
\item {\em Interested Message Receive Rate}. For each message, we compute the ratio of the number of subscribers in the message's topic that received the message and the total number of subscribers in the message's topic and plot the average over the 1000 messages for each epoch. 
\item {\em Average Propagation Delay}. We measure the time between when a message is published and when the message is received by a subscriber and average the measurements.
We ignore subscribers that do not receive the message in the averaging. 
\item {\em Average Neighbor Score}. Nodes score their neighbors at the end of each epoch, we average the scores among all the nodes.
\item {\em Score Distribution}. We collect the scores from all the nodes, sort them and color them by the epochs.
\end{enumerate}

\subsection{Results}
\begin{figure}[!t]
\centering
  \includegraphics[width=0.5\textwidth]{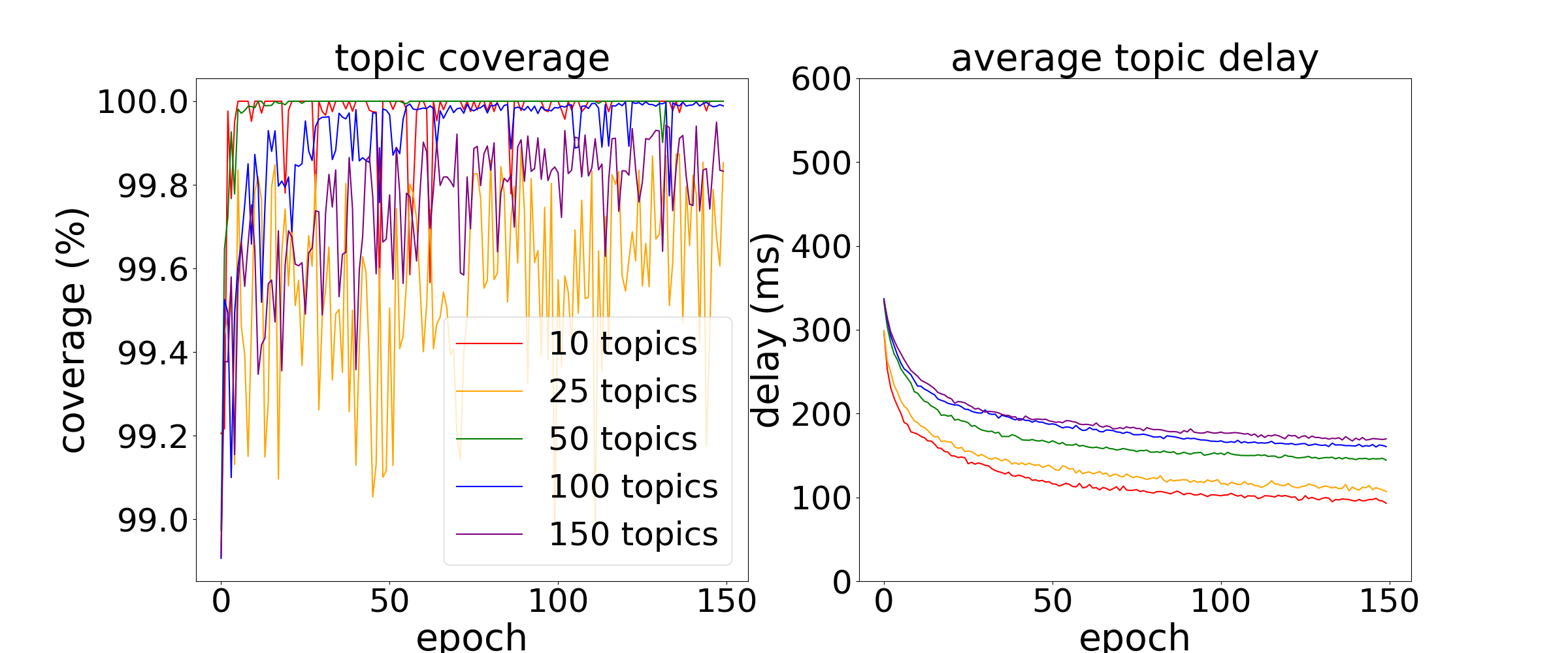}
  \caption{Network performance under various interested topics}
  \label{fig:topic number}
\end{figure}

\subsubsection{Number of topics}
\label{s: fewtopics}
We first consider the 1000-node network within a unit square. 
We run the network with 10, 25, 50, 100, 150 topics with an interest rate of 40\% for each topic.
As plot in Figure~\ref{fig:topic number}, networks with various topics converge after around 50 epoches.
They reach around 100\% interesting message coverage and 50\% lower latency to receive their interesting topic. .

\begin{figure*}[!t]
\centering
  \includegraphics[width=0.85\textwidth]{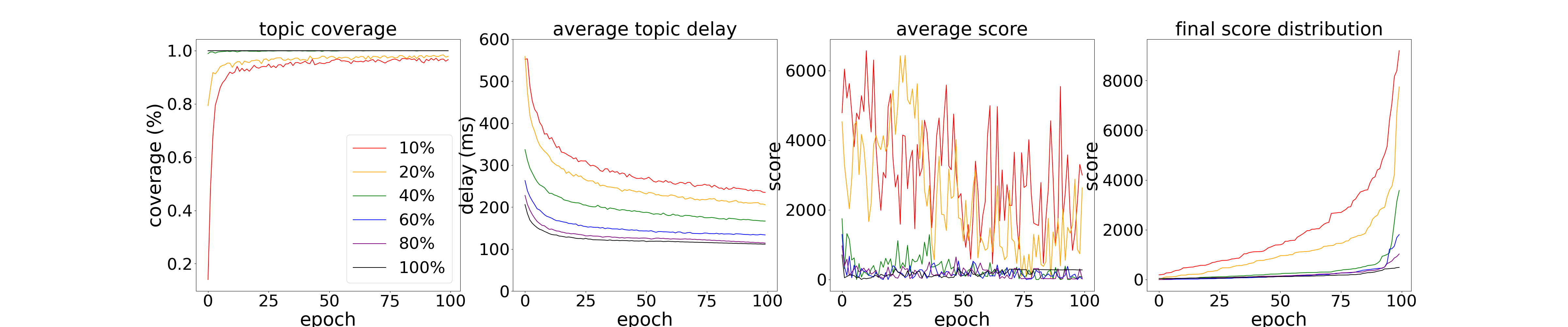}
  \caption{Network performance under various topic interest rate}
  \label{fig:interest rate}
\end{figure*}
\smallskip 
\subsubsection{Interest rate}
\label{s: manytopics}
Next we consider the network with 100 topics but with different interest rate of 10\%, 20\%, 40\%, 60\%, 80\% and 100\% per topic.
An interest rate of 100\% means that nodes favor all the topics, which is similar to the single topic network.
Network also converges after a50 epoches. 
With around 100\% topic coverage, topics with lower interest rate get more advantage, where the network with 10\% interest has improvement more than 60\%.

\begin{figure}[!t]
    \centering
  \includegraphics[width=0.5\textwidth]{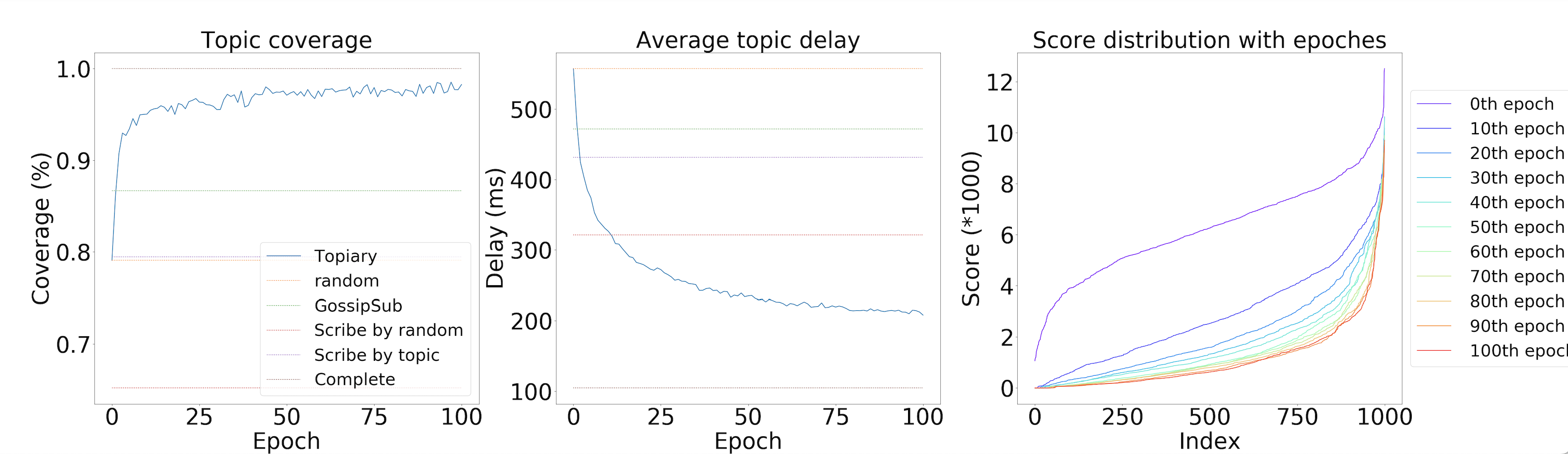}
  \caption{Network performance among different topoligies}
  \label{fig:comparison}
\end{figure}
\subsubsection{Topology comparison}
\label{s: comparison}
We show the Topiary's performance in the above paragraphs, here we focus on 100 topics with interest rate of 20\% to compare with the other base line topologies.
We evaluate the performance by the topic coverage and topic propagation latency as plot in Figure~\ref{fig:comparison}.
Topiary starts with random state, so that the random algorithm is the performance of Topiary at 1st epoch.
Scribe groups nodes, and builds tree structure based on the nodes' id.
Here we implement Scribe in 2 methods, to groups nodes randomly or group nodes by topics.
Randomly groups has 40 groups, it has a low delay but poor connectivity.
Topic based grouping has 100 groups, it has a lower delay and lower coverage compared with the others.
Gossipsub also cannot support a large number of topics without violating degree bound constraints.
In Topiary, after around 10 epochs of  learning, messages reach more than 90\% of the interested nodes. 
After 50 epochs of training, Topiary converges with fewer improvements in receive rate and average propagation delay.
The corresponding average score by the score function also converges.
We can find the converged state has around 250ms average delay and 98\% receive rate for the 10\% interest rate case, 210ms average delay and 98\% receive rate with  20\% interest rate, and 180ms average delay in the 30\% interest rate case.
Thus, Topiary can find efficient topologies even under a large number of topics. 

\subsubsection{Parameter sensitivity}
\label{s: parameterstudy}
\begin{figure}[!t]
    \centering
  \includegraphics[width=0.5\textwidth]{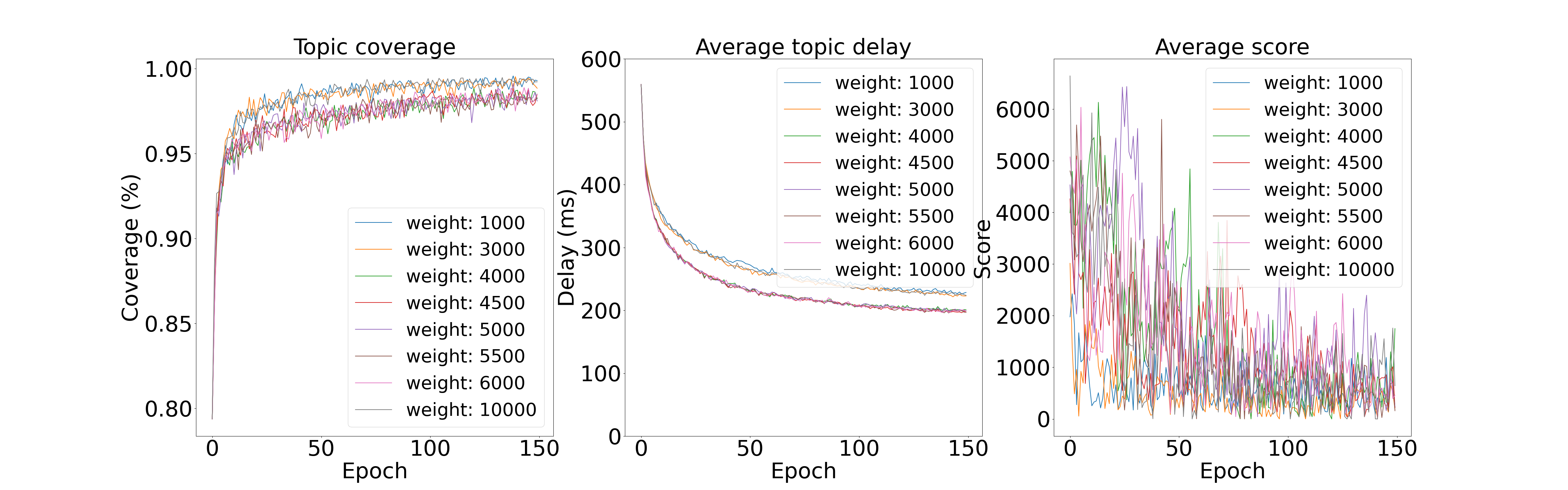}
  \caption{Network performance by different $w_d$ in Equation~\eqref{eq:overall score}.}
  \label{fig:parameter_senstivity}
\end{figure}

We study the sensitivity of weight parameter $w_d$ in Equation~\eqref{eq:overall score}.
We normalize the equation by $w_c=1$, and set $w_w=0$ in the experiments with high topic coverage and tune $w_d$ from 1000 ti 10,000.
Figure~\ref{fig:parameter_senstivity} shows that there's a trade-off between topic coverage and topic broadcasting delay with the weight on delay in the score function.

\subsubsection{Real world network}
\label{s: readworld}
\begin{figure}[!t]
\centering
  \includegraphics[width=0.5\textwidth]{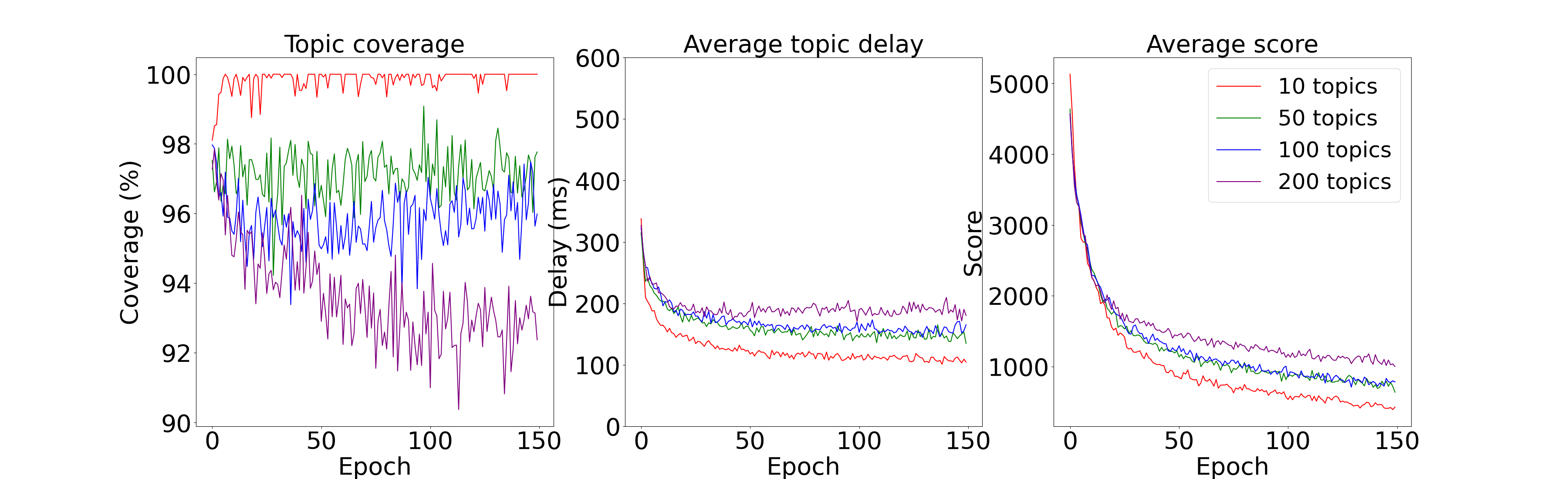}
  \caption{Network performance in 246-node network}
  \label{fig:246-node_performance}
\end{figure}

\begin{figure}[!t]
\centering
  \includegraphics[width=0.5\textwidth]{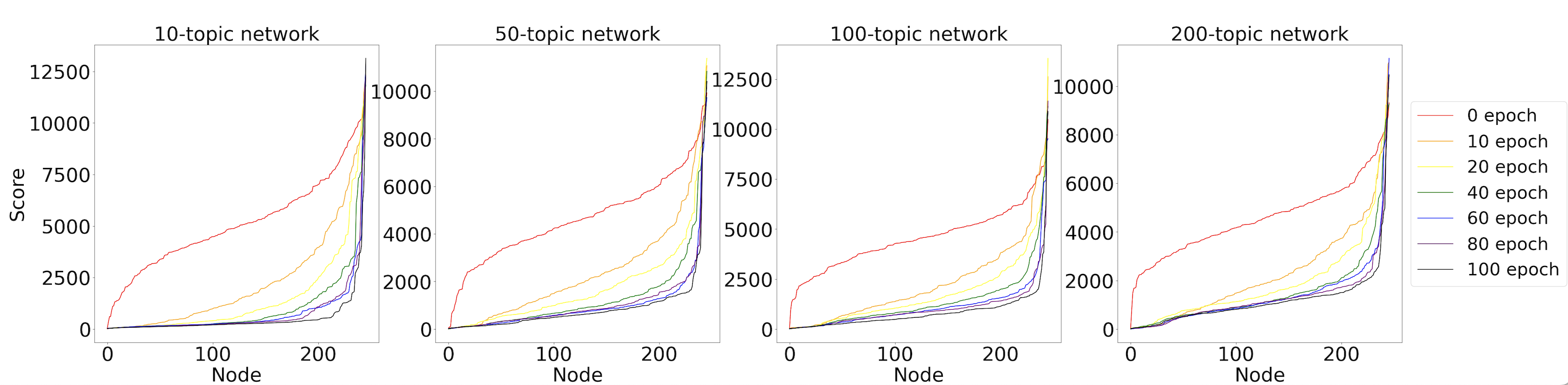}
  \caption{Score distribution with epoches in 246-node network}
  \label{fig:246-node_score_distribution}
\end{figure}

Topiary in the 246 node real world network converges after 50 epochs with around 50\% improvements compared with the random topology and close to 100\% interested message coverage to the whole network with the same setting of topic amount and topic sizes in Figure~\ref{fig:246-node_performance} and Figure~\ref{fig:246-node_score_distribution}.

\subsection{Experiments on 246-node network with reality data}

\subsection{Security}
\label{s: security}

\begin{figure}[!t]
  \includegraphics[width=0.5\textwidth]{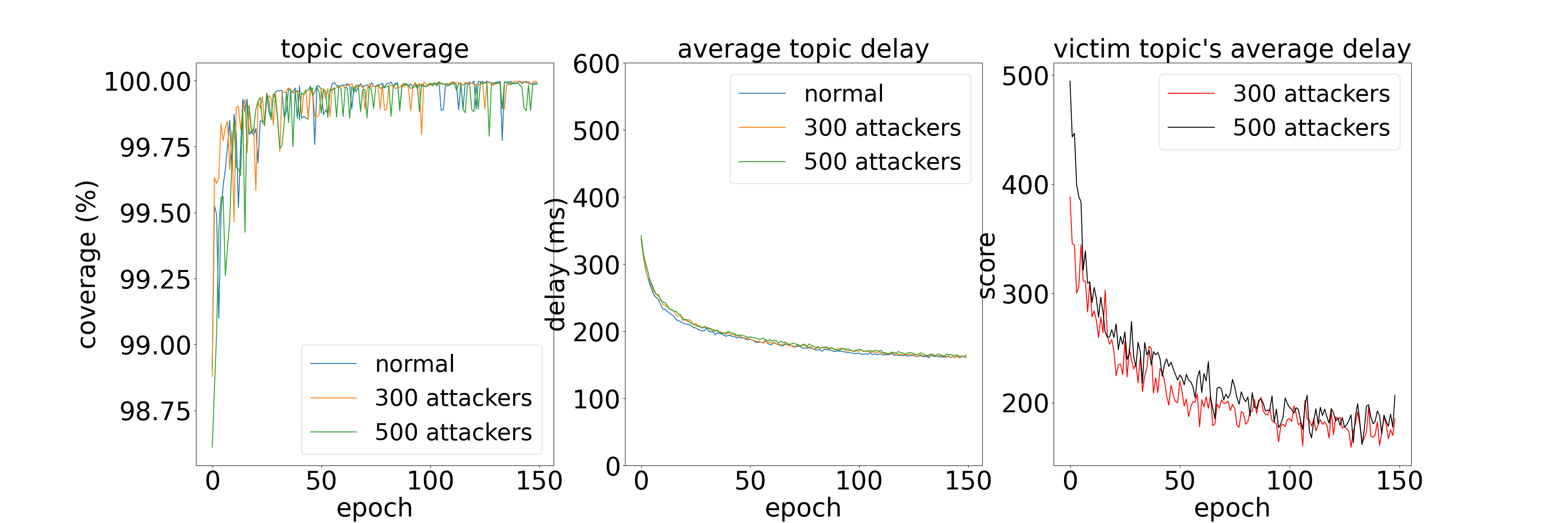}
  \caption{Network performance after attacker by 300/500 topic-based attackers}
  \label{fig:topic attack}
\end{figure}
\subsubsection{Topic based attack}
\label{s: topic attack}
We first focuses on topic-based attacks within the network. These attackers participate in the pub/sub network by professing interest in the victim topic but abstain from engaging in message forwarding under that specific topic.
We introduce two sets of attackers, totaling 300 or 500 individuals, within a network comprising 1000 nodes, all expressing interest in 100 topics at a rate of 40\%. Figure~\ref{fig:topic attack} illustrates the consequential impact of such a substantial number of attackers on both topic coverage and topic delay.
The third column of the figure portrays the average delay associated with the victim topic, eventually aligning with the average delay across all 100 topics. Through Topiary's scoring mechanism applied to neighbor subsets, nodes can discern and identify attackers displaying negative actions.

\subsubsection{Eclipse Attack}
\label{s: Eclipse Attack}
To simulate an Eclipse attack~\cite{wust2016ethereum}, we introduce 300 attacker nodes to the network among 1000-node network with the same topic interest and complete connections among attacker.

\begin{figure}[!t]
  \includegraphics[width=0.45\textwidth]{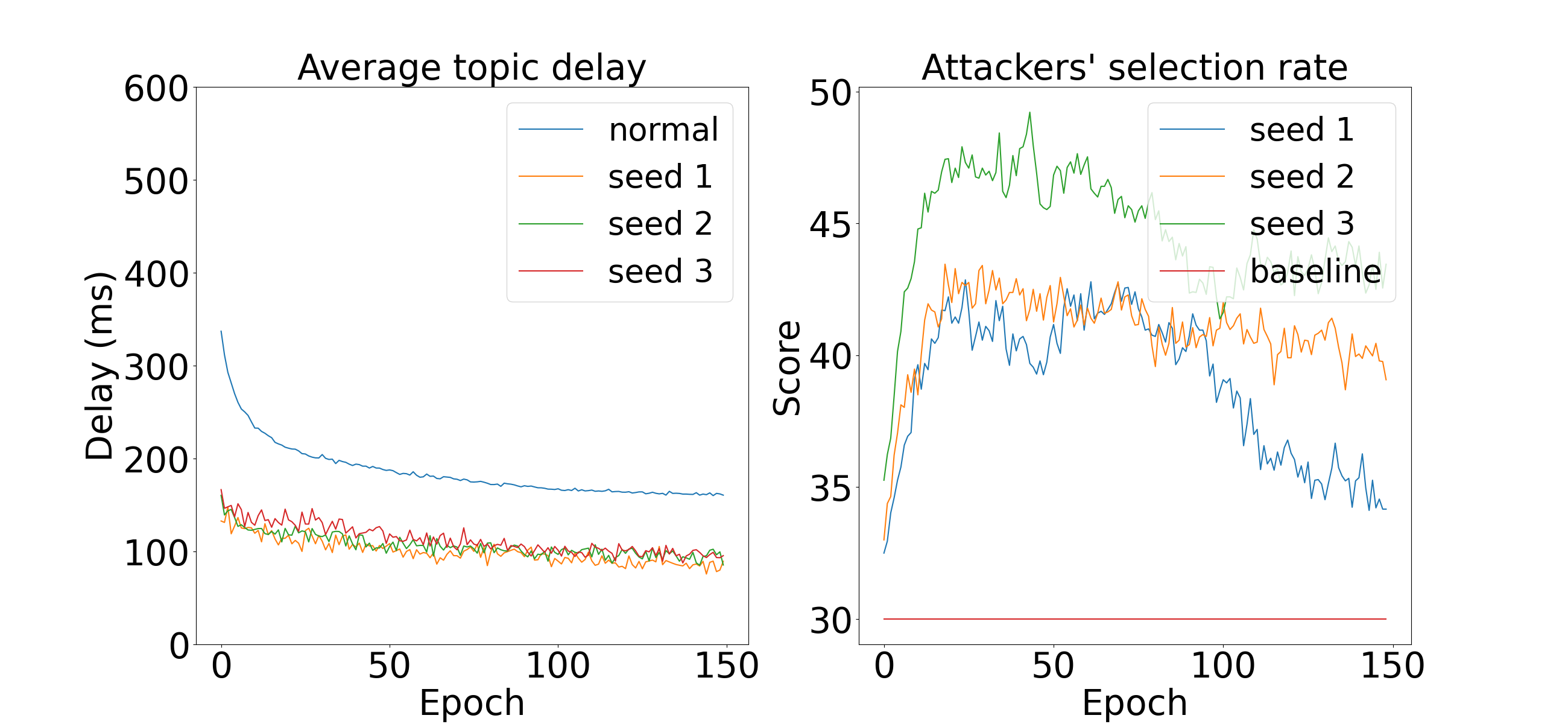}
  \caption{Network performance after introducing 300 eclipse attackers}
  \label{fig:eclipse attack}
\end{figure}

Figure~\ref{fig:eclipse attack} shows that the network can get advantage from the complete 300-attacker connections.
Messages get lower delay in propagating.
The 2nd figure shows the percentile of the 300 attackers to be selected as the outgoing neighbors from the other 700 nodes.
The attackers are more favored than the other nodes, but they can hardly occupy all the honest nodes' outgoing degree, which prevent them to conduct future attacks to isolate some honest nodes.

Our results show that Topiary can protect the pub/sub network from topic based attack, nodes can figure out the negative nodes and updated their next epoch neighbors to optimized their neighbor selection.
Topiary also guides nodes to distribute their pub/sub requirements from the combination of neighbors, the potential eclipse attackers can hardly occupy all their neighbor connections.

\section{Related Work}
\label{s: related}
Ethereum implements pub/sub in its topic advertisement subsystem~\cite{ethNodeDiscovery} using a node discovery protocol. 
This protocol facilitates an easy approach for nodes to advertise themselves on the network and assists in exploring the network to access required services. 
A node can become discoverable and act as a publisher under a specific topic. 
When other nodes in the network accept such services, they gain knowledge of the new publisher and can share the same search results with their unknown neighbors, aiding in broadcasting topic information. 
They maintain a node storage system based on topic interests and within a certain proximity of their Kademila addresses~\cite{zhang2023Kadabra}. 
Subscribers can locate publishers offering specific topic services through this system.

IPFS merges a pub/sub implementation at 2017~\cite{ipfspubsub}. 
The pub/sub implementation makes IPFS faster for large-scale networks such as datacenters, local area networks, and large p2p applications.
IPFS search is a search engine for the interplanetary filesystem~\cite{ipfssearch}.
It offers a decentralized file storing and sharing system that splits files into smaller chunks with a content identifier.
Within the IPFS network, nodes store these chunks and act as publishers for their respective chunk topics. 
When nodes seek an entire file, they query their network neighbors to retrieve the required chunks and establish connections to nodes holding other chunks within the same topic.
Once nodes obtain the desired chunks, they assume the role of publishers for that topic, enabling subsequent searches from other nodes looking for the same topic.

Streamr~\cite{savolainen2020streamr} orchestrates the network topology with the assistance of trackers to maintain low and predictable latency. These trackers are assigned to aid nodes in discovering peers and joining the overlay.
The role of the tracker involves managing the topology of streams by guiding nodes on connection and disconnection instructions, facilitating the calculation of event propagation delay within the current topology using Dijkstra's algorithm.
Streamr initially establishes a well-connected network state by pairing nodes between groups. 
Subsequently, it designs algorithms to maintain network stability amidst minor network alterations.

There are some pub/sub algorithms that aim to help the nodes to choose well-behaved neighbors and optimize the network topology.
They are designed for various network environments like the network size, topic size, and node degree bound.
We have some brief descriptions of their topology strategies for them below with comparisons of Scribe~\cite{1038579} and  GossipSub~\cite{vyzovitis2020gossipsub} in Section~\ref{s: evaluation} with the same network setting.

Scribe~\cite{1038579} is designed to create a multicast infrastructure capable of supporting a large number of groups with varying sizes. The grouping strategy can rely on randomness or similarities among nodes~\cite{zhuang2001bayeux,ratnasamy2001application,rahimian2011vitis}, such as nodes' network states or topics, as discussed in this paper. However, complex topic subscriptions can make grouping nodes challenging. Grouping nodes without topic information or with specific topic information might not yield a topology suitable for the entire pub/sub network.

Systems like Tera~\cite{baldoni2007tera}, Spidercast~\cite{chockler2007spidercast}, Poldercast~\cite{setty2012poldercast}, and GossipSub~\cite{vyzovitis2020gossipsub} organize groups based on topic information, fostering stronger connections within each group. GossipSub, for instance, distributes its degree among meshes based on the interested topics, leveraging robust connections among nodes with the same topic while ensuring broader coverage across multiple topics with the remaining degree. Nevertheless, these systems may struggle to handle networks requiring numerous topic types due to limitations on the degree bound.

\section{Conclusion}
\label{s: conclusion}
We've explored the challenge of topology design in optimizing neighbor selection within pub/sub networks. By employing Topiary, nodes can establish a scoring function for their neighbors, enabling them to judiciously select some neighbors while exploring alternative connections with the remaining ones.
The findings reveal that nodes achieve a notably improved network connection state, ensuring the delivery of at least 98\% of the messages related to desired topics. Moreover, this optimized setup demonstrates a 50\% reduction in propagation delay compared to a randomly constructed network, even in high-demand scenarios for specific topics, surpassing the network degree.
This enhanced network state is typically attained after approximately 50 epochs. Subsequently, the network consistently maintains a robust connection state, boasting extensive coverage of desired messages and consistently low propagation delays.

We additionally conducted tests involving Topic-based attackers and Eclipse attackers within the network.
Utilizing our neighbor scoring method, nodes evaluate neighboring nodes based on their subset's performance. Consequently, Sybil attackers employing prolonged node processing delays are promptly identified and discarded due to their underperformance.
The incorporation of random exploration in neighbor switching ensures that all nodes focusing on the target topic have an equal probability of selection. This strategic approach significantly mitigates the efficacy of Eclipse attackers in obstructing a node, as each node maintains a probability of establishing connections with other nodes, thus reducing the likelihood of successful blocking.
  
\bibliographystyle{IEEEtran}
\bibliography{paper}

\begin{thebibliography}{10}
\providecommand{\url}[1]{#1}
\csname url@samestyle\endcsname
\providecommand{\newblock}{\relax}
\providecommand{\bibinfo}[2]{#2}
\providecommand{\BIBentrySTDinterwordspacing}{\spaceskip=0pt\relax}
\providecommand{\BIBentryALTinterwordstretchfactor}{4}
\providecommand{\BIBentryALTinterwordspacing}{\spaceskip=\fontdimen2\font plus
\BIBentryALTinterwordstretchfactor\fontdimen3\font minus
  \fontdimen4\font\relax}
\providecommand{\BIBforeignlanguage}[2]{{%
\expandafter\ifx\csname l@#1\endcsname\relax
\typeout{** WARNING: IEEEtran.bst: No hyphenation pattern has been}%
\typeout{** loaded for the language `#1'. Using the pattern for}%
\typeout{** the default language instead.}%
\else
\language=\csname l@#1\endcsname
\fi
#2}}
\providecommand{\BIBdecl}{\relax}
\BIBdecl

\bibitem{savolainen2020streamr}
P.~SAVOLAINEN, S.~JUSLENIUS, E.~ANDREWS, M.~POKROVSKII, S.~TARKOMA, and
  H.~PIHKALA, ``The streamr network: Performance and scalability,'' \emph{url:
  https://streamrpublic. s3. amazonaws.
  com/streamr-network-scalability-whitepaper-2020-08-20. pdf}.

\bibitem{al2018fraud}
M.~Al-Bassam, A.~Sonnino, and V.~Buterin, ``Fraud and data availability proofs:
  Maximising light client security and scaling blockchains with dishonest
  majorities,'' \emph{arXiv preprint arXiv:1809.09044}, 2018.

\bibitem{vyzovitis2020gossipsub}
D.~Vyzovitis, Y.~Napora, D.~McCormick, D.~Dias, and Y.~Psaras, ``Gossipsub:
  Attack-resilient message propagation in the filecoin and eth2. 0 networks,''
  \emph{arXiv preprint arXiv:2007.02754}, 2020.

\bibitem{mao2020perigee}
Y.~Mao, S.~Deb, S.~B. Venkatakrishnan, S.~Kannan, and K.~Srinivasan, ``Perigee:
  Efficient peer-to-peer network design for blockchains,'' in \emph{Proceedings
  of the 39th Symposium on Principles of Distributed Computing}, 2020, pp.
  428--437.

\bibitem{bowengoldfish}
B.~Xue, Y.~Mao, , S.~B. Venkatakrishnan, and S.~Kannan, ``Goldfish: Peer
  selection using matrix completion in unstructured p2p network,'' in
  \emph{IEEE International Conference on Blockchain and Cryptocurrency}, 2023.

\bibitem{auer2002using}
P.~Auer, ``Using confidence bounds for exploitation-exploration trade-offs,''
  \emph{Journal of Machine Learning Research}, vol.~3, no. Nov, pp. 397--422,
  2002.

\bibitem{babel2022strategic}
K.~Babel and L.~Baker, ``Strategic peer selection using transaction value and
  latency,'' in \emph{Proceedings of the 2022 ACM CCS Workshop on Decentralized
  Finance and Security}, 2022, pp. 9--14.

\bibitem{math11234741}
\BIBentryALTinterwordspacing
Y.~Mao and S.~B. Venkatakrishnan, ``Less is more: Understanding network bias in
  proof-of-work blockchains,'' \emph{Mathematics}, vol.~11, no.~23, 2023.
  [Online]. Available: \url{https://www.mdpi.com/2227-7390/11/23/4741}
\BIBentrySTDinterwordspacing

\bibitem{gai2010learning}
Y.~Gai, B.~Krishnamachari, and R.~Jain, ``Learning multiuser channel
  allocations in cognitive radio networks: A combinatorial multi-armed bandit
  formulation,'' in \emph{2010 IEEE Symposium on New Frontiers in Dynamic
  Spectrum (DySPAN)}.\hskip 1em plus 0.5em minus 0.4em\relax IEEE, 2010, pp.
  1--9.

\bibitem{kveton2015tight}
B.~Kveton, Z.~Wen, A.~Ashkan, and C.~Szepesvari, ``Tight regret bounds for
  stochastic combinatorial semi-bandits,'' in \emph{Artificial Intelligence and
  Statistics}.\hskip 1em plus 0.5em minus 0.4em\relax PMLR, 2015, pp. 535--543.

\bibitem{chen2013combinatorial}
W.~Chen, Y.~Wang, and Y.~Yuan, ``Combinatorial multi-armed bandit: General
  framework and applications,'' in \emph{International conference on machine
  learning}.\hskip 1em plus 0.5em minus 0.4em\relax PMLR, 2013, pp. 151--159.

\bibitem{dong2023graph}
G.~Dong, M.~Tang, Z.~Wang, J.~Gao, S.~Guo, L.~Cai, R.~Gutierrez, B.~Campbel,
  L.~E. Barnes, and M.~Boukhechba, ``Graph neural networks in iot: a survey,''
  \emph{ACM Transactions on Sensor Networks}, vol.~19, no.~2, pp. 1--50, 2023.

\bibitem{wondernetwork}
``Global ping statistics,'' \url{https://wondernetwork.com/pings}.

\bibitem{wust2016ethereum}
K.~W{\"u}st and A.~Gervais, ``Ethereum eclipse attacks,'' ETH Zurich, Tech.
  Rep., 2016.

\bibitem{ethNodeDiscovery}
``Ethereum node discovery protocol v5,''
  \url{https://github.com/ethereum/devp2p/blob/master/discv5/discv5.md}.

\bibitem{zhang2023Kadabra}
S.~B.~V. Yunqi~Zhang, ``Kadabra: Adapting kademlia for the decentralized web,''
  \emph{arXiv preprint arXiv:2210.12858}, 2023.

\bibitem{ipfspubsub}
``Take a look at pubsub on ipfs,'' \url{https://blog.ipfs.tech/25-pubsub/}.

\bibitem{ipfssearch}
``Ipfs search,'' \url{https://ipfs-search.com/#/}.

\bibitem{1038579}
M.~Castro, P.~Druschel, A.-M. Kermarrec, and A.~Rowstron, ``Scribe: a
  large-scale and decentralized application-level multicast infrastructure,''
  \emph{IEEE Journal on Selected Areas in Communications}, vol.~20, no.~8, pp.
  1489--1499, 2002.

\bibitem{zhuang2001bayeux}
S.~Q. Zhuang, B.~Y. Zhao, A.~D. Joseph, R.~H. Katz, and J.~D. Kubiatowicz,
  ``Bayeux: An architecture for scalable and fault-tolerant wide-area data
  dissemination,'' in \emph{Proceedings of the 11th international workshop on
  Network and operating systems support for digital audio and video}, 2001, pp.
  11--20.

\bibitem{ratnasamy2001application}
S.~Ratnasamy, M.~Handley, R.~Karp, and S.~Shenker, ``Application-level
  multicast using content-addressable networks,'' in \emph{Networked Group
  Communication: Third International COST264 Workshop, NGC 2001 London, UK,
  November 7--9, 2001 Proceedings 3}.\hskip 1em plus 0.5em minus 0.4em\relax
  Springer, 2001, pp. 14--29.

\bibitem{rahimian2011vitis}
F.~Rahimian, S.~Girdzijauskas, A.~H. Payberah, and S.~Haridi, ``Vitis: A
  gossip-based hybrid overlay for internet-scale publish/subscribe enabling
  rendezvous routing in unstructured overlay networks,'' in \emph{2011 IEEE
  International Parallel \& Distributed Processing Symposium}.\hskip 1em plus
  0.5em minus 0.4em\relax IEEE, 2011, pp. 746--757.

\bibitem{baldoni2007tera}
R.~Baldoni, R.~Beraldi, V.~Quema, L.~Querzoni, and S.~Tucci-Piergiovanni,
  ``Tera: topic-based event routing for peer-to-peer architectures,'' in
  \emph{Proceedings of the 2007 inaugural international conference on
  Distributed event-based systems}, 2007, pp. 2--13.

\bibitem{chockler2007spidercast}
G.~Chockler, R.~Melamed, Y.~Tock, and R.~Vitenberg, ``Spidercast: a scalable
  interest-aware overlay for topic-based pub/sub communication,'' in
  \emph{Proceedings of the 2007 inaugural international conference on
  Distributed event-based systems}, 2007, pp. 14--25.

\bibitem{setty2012poldercast}
V.~Setty, M.~Van~Steen, R.~Vitenberg, and S.~Voulgaris, ``Poldercast: Fast,
  robust, and scalable architecture for p2p topic-based pub/sub,'' in
  \emph{Middleware 2012: ACM/IFIP/USENIX 13th International Middleware
  Conference, Montreal, QC, Canada, December 3-7, 2012. Proceedings 13}.\hskip
  1em plus 0.5em minus 0.4em\relax Springer, 2012, pp. 271--291.

\end{thebibliography}

\end{document}